# On Social Machines for Algorithmic Regulation

Nello Cristianini and Teresa Scantamburlo

University of Bristol

**Abstract**

Autonomous mechanisms have been proposed to regulate certain aspects of society and are already being used to regulate business organisations. We take seriously recent proposals for algorithmic regulation of society, and we identify the existing technologies that can be used to implement them, most of them originally introduced in business contexts. We build on the notion of 'social machine' and we connect it to various ongoing trends and ideas, including crowdsourced task-work, social compiler, mechanism design, reputation management systems, and social scoring. After showing how all the building blocks of algorithmic regulation are already well in place, we discuss possible implications for human autonomy and social order. The main contribution of this paper is to identify convergent social and technical trends that are leading towards social regulation by algorithms, and to discuss the possible social, political, and ethical consequences of taking this path.

## 1 - Introduction

A recent article by historian Yuval Harari argues that "the conflict between democracy and dictatorship" is not one "between different ethical systems, but actually between data-processing systems." (Harari, 2018). While in the 20th century liberal democracies benefited from their decentralised approach to decision-making – distributing information and power among many people proved to be more effective than concentrating them in one place – in the 21st century recent progress in Artificial Intelligence (AI) has turned this around by enabling intelligent systems to process large volumes of information centrally (Harari, 2018). According to Harari, this shift would suggest that democratic ideals, such as equality and liberty, are "more fragile than we believe" (Harari 2018). Thus, rather than being "self-evident" or "irreversible" (Harari 2018), these ideals may change subtly, in a way that we do not expect, and we do not want.

This article is concerned with the notion of "social regulation", by which we mean the activity of governing a society, encouraging certain outcomes over others, steering the behaviour of a community. While this activity has traditionally been performed by a complex of explicit and implicit rules, enacted by an authority, or elicited by social interactions, we are interested in how modern AI technology interacts with it.

The problem of effectively governing a country of hundreds of millions of citizens has been debated in policy circles for a long time (Heaven, 2017), and there are proposals to turn to digital technology (Larson, 2018). In this article, we will use the expression "algorithmic regulation" to refer to the use of algorithmic methods for social regulation or governance[1].

As we debate new ways to apply intelligent technologies to governance, we cannot ignore that modern web companies manage numbers of users larger than that of most countries (Constine, 2017), nor can we ignore the problems posed by the deployment of AI in restricted domains such as personalised news delivery and enforcement decisions (e.g. Mittelstadt, 2016; Burr et al, 2018, Scantamburlo et al, 2019). What could be the effect of deploying AI systems at the centre of a growing infrastructure of linked-data for governance of a whole society? To answer this question, we take seriously the proposal made by Tim O'Reilly for *algorithmic regulation* of society, examine its risks, and its relation to experiments currently under way in different countries and the private sector.

In 2013, the Silicon Valley investor, publisher, and futurist Tim O'Reilly proposed that society can be more effectively regulated by using feedback loops, rather than top-down law enforcement. His example was the way in which ride-sharing apps (e.g. Uber) can regulate the behaviour of both drivers and passengers, both by leveraging a range of sensor data and by maintaining a reputation management system, instead of using top-down rules and inspection (O'Reilly, 2013). O'Reilly also drew general conclusions for social governance, suggesting that data sources, combined with a reputation management system can do a better job than any amount of government regulation (O'Reilly, 2013:293). Taxis can "increase their availability even in less frequented locations" and, more importantly, by asking passengers to rate their drivers, the quality of services can improve automatically: "drivers who provide poor service are eliminated" (O'Reilly, 2013, 293). The central elements indicated by O'Reilly for his notion of algorithmic regulation are: clear outcomes, real-time measurements indicating whether those outcomes have been achieved, and adjustments of the rules based on such measurements (O'Reilly, 2013, 290). Of course, these are also the central requirements of feedback-control systems in engineering, as will be discussed below.

---

[1] Other definitions have been proposed by Yeung (2017) and Hildebrandt (2018). Our discussion connects also to Gillespie (2017) but it extends beyond the problem of how social media platforms curate contents and police the activity of their users by algorithms.

Similar proposals have recently gained importance, since the government of China announced the creation of a national scoring system for its citizens that can ultimately be used to administer positive and negative incentives, either in the form of these citizens being added to or removed from certain lists, and in some cases in the form of points assigned to citizens based on their behaviour (Creemers, 2018). Bad behaviours (also called "trust-breaking acts"), such as tax evasion or refusal to carry out statutory duties, contribute new entrants in a blacklist and this affects both individuals and organisations by imposing penalties and limiting access to government programmes, high-speed trains, hotels or purchases, etc.[2]

While governments and public organisations are moving forward, the private sector went ahead with the employment of pervasive tracking and scoring mechanisms for the enforcement of desired behaviours. In this way, eBay allows buyers and sellers to rate one another to ensure successful trading, health insurance companies[3] use trackers and personalised incentives to help their customers "live longer" (Sullivan, 2018), and many apps for housekeeping activities, like Handy, work in a Uber-style fashion to offer efficient and affordable services (Ticona et al., 2018). Likewise, Amazon fulfilment centres constantly monitor and analyse their employees to meet certain performance standards (Bernton and Kelleher, 2012) and, as we recently discovered, Facebook scores its users' trustworthiness when they flag a post as fake (Dwoskin, 2018).

---

[2] As well as Creemers (2018) see also the following document: "State Council Guiding Opinions concerning Establishing and Perfecting Incentives for Promise-keeping and Joint Punishment Systems for Trust-Breaking, and Accelerating the Construction of Social Sincerity." State Council. 30th May 2016. Available online: https://chinacopyrightandmedia.wordpress.com/2016/05/30/state-council-guiding-opinions-concerning-establishing-and-perfecting-incentives-for-promise-keeping-and-joint-punishment-systems-for-trust-breaking-and-accelerating-the-construction-of-social-sincer/.

[3] For example, Vitality programme (https://www.vitalitygroup.com/) provides customers with a smartphone app to set up personalised goals, carry out and monitor activities (e.g. buying healthy food, doing physical exercise, spinning wheel, etc). The application includes the use of personal data, gamified elements and scoring mechanism (Sullivan, 2018).

China's social credit score represents a paradigmatic example, but parts of the anatomy of the Chinese system – individual ID, linked data, automated interventions – (Creemers, 2018) are in place, or under way, also in other countries such as Singapore and Estonia. In most of these attempts there are plans to integrate services which may involve citizens, associations or business activities, from voting to education, taxes and justice, in one single infrastructure that can run processes by itself (filing taxes, reviewing medical records, checking eligibility for voting, etc.) and influence collective behaviour[4]

To analyse the main technical and sociological underpinnings of O'Reilly's proposal we will review the various components that would be needed to implement that type of algorithmic regulation of a society, and how these components are being used at the moment. In doing so, we encounter the notions of reputation management, social machines, and autonomous agents, among others, and we propose a unified framework that can help us understand how close we already are to O'Reilly's vision of a society regulated by algorithms, and what might be some of its consequences.

In this article, motivated by O'Reilly's intuition, we relate the notion of algorithmic regulation to the use of digital systems to monitor citizens and give them automatic and personalised incentives with the purpose of influencing their behaviour. The question we want to pose concerns the implications of using algorithmic regulation technologies to govern a society.

To summarise our main findings: social machines are a stable technology that is used, among other things, to create reputation management systems; the principles and details of their design

---

[4] For example, albeit different, both China' Social Credit Score and Estonia' e-government aim at promoting certain behaviours within society. So, the Chinese ambition is to stimulate sincerity and trust (Creemers, 2018), while the Estonia's goal is to foster transparency and "to make it impossible to do bad things" (Keen, 2016)

are extremely important for considerations relative to user autonomy (mechanism design and crowdsourcing being parts of the discussion); once created, these machines can act as autonomous agents, and act in a way to maximise their utility (whose alignment with collective utility should not be taken for granted) generating important technical and social challenges relative to their stability, alignment, transparency and fairness. Feedback loops and control are other key features of such systems and may interfere with fundamental aspects of our society such as people's autonomy, social order and the exercise of power.

This article is organised as follows. In section 2 we will describe the notion of social machines, how they can be implemented and how they can be used to implement autonomous agents. In section 3 we will describe reputation (management) systems, how they relate to social scoring and how they can be part of a regulatory feedback loop. In section 4 we will put these two ideas together to describe how we already have social regulation via distributed reputation scoring, all mediated by a single algorithm, in limited domains. As these methods are currently used in the private sector, we briefly look at what can be learnt from that experience, before discussing current efforts to deploy them to regulate entire societies. In section 5 we will discuss some considerations about the technical and social consequences of this proposed turn and summarise our findings in the conclusions.

## 2 - Social Machines as Autonomous Systems

The essence of algorithmic regulation proposals lies in the automatic delivery of positive and negative incentives (e.g., discounts or fines) to individuals for specific actions, behaviours or performance. These incentives are aimed at individuals, and therefore require ways to collect individual information, resulting either from measurements (e.g., time required to perform a delivery) or from social feedback (e.g., customer satisfaction rating, as proposed in O'Reilly, 2013).

In this sense, their implementation would not be very different than for existing forms of personalisation (e.g., personalised recommendations or ads, or credit scores), and would typically require a way to identify individuals, and gather and store individual information, including some sort of a score. The citizens would be expected to adjust their behaviour in a way to improve their score, if it is linked to benefits, opportunities, or even just social recognition.

While some of the required individual information would be directly observed and measured (e.g. geolocation data, payments, etc.), a part of it would be the product of human judgment (as in the car-sharing example). In order to understand the functioning of this class of systems (e.g., reputation systems) that bring together algorithms and humans we will employ the more general notion of social machines.

2.1 – Social Machines

A machine is a system, or apparatus, formed by several parts, each with a definite function, which interact together in a specific way, so as to perform a particular task. There is no limitation to the technical substrate of these parts, e.g., they can include hydraulic, electric, or mechanical parts, among others. A social machine is a special type of machine where some of the components, performing some subtask, are formed by humans (whom we call participants).

The moving assembly line is an example of a social machine. An assembly line is formed by a set of workstations where the same operations are always performed in a consistent way, and various parts are added to a product, as it moves through the line. Some of the operations are performed by machines, and others by people, in a highly coordinated and systematic fashion. So long as all the operations are performed in the same time and way, it does not matter who performs them. Human participants are typically used for operations that cannot be easily automated, but act in very structured manner, and do not control the overall process. They are in fact parts of a machine and do not need to be aware of the overall results of their actions in order to do their job.

A bureaucracy is another, classic form of social machine[5]. For example, take a national post office, a bank or an airline, they all share a set of structural characteristics: e.g. functions and roles reflect a hierarchy, tasks are divided among workers and performed routinely, the input and the output of the tasks are standardised, the workflow and the coordination among the workers are specified by rules and communications are performed via structured forms. Even though many tasks are performed by humans, each participant has limited autonomy and is not in the position to determine the behaviour of the overall machine, maybe not even be aware of it.

---

[5] Max Weber (1978) was the first to acknowledge this and considered bureaucracy as the most effective form of organisation: "The decisive reason for the advance of bureaucratic organization has always been its purely technical superiority over any other form of organization. The fully developed bureaucratic apparatus compares with other organizations exactly as does the machine with the non-mechanical modes of production." (Weber, 1978, 973)

Drawing on a rich literature (e.g. see Berners-Lee and Fischetti, 1999; Smart and Shadbolt, 2014) we define a social machine as a machine where human participants and technical artefacts (e.g. a car, a piece of software, a robot) interact with one another to perform a task that would be hardly achievable by any single part[6]. Mechanisms incorporating 'participants' extend across domains and include: assembly lines, bureaucracies, auctions, markets, voting schemes, product delivery services, games, peer production, crowdsourcing, etc.

Even though social machines have existed for a long time, they have been formalised only recently in the context of web technologies (Berners-Lee and Fischetti, 1999) and include various ways in which communities are organised by a web-based infrastructure[7]. Online crowdsourcing services, such as Amazon's Mechanical Turk, operate as modern assembly lines, where each participant performs a well specified task, that might be difficult to automate, and does not need to know the overall goals of the machine. For example, participants can be asked to tag faces in photos by gender, to annotate images or articles, to type handwritten words, to answer questions and so forth.

Participants in crowdsourcing not only might not know the purpose of the machine, they might not even know its boundaries, i.e. what else is part of it. In other words, they are not in the position to control the machine's overall behaviour. The use of web infrastructures to coordinate the behaviour of participants has proven to be a very effective way to organise social machines that may reach sizes of millions of participants.

---

[6] Note that the meaning of "social machine" overlaps many popular abstractions such as "collective intelligence", "distributed cognition", "wisdom of crowd", "social computing", "social computation."
[7] Note that when referred to web infrastructures, the notion of social machine incorporates that of online platform developed in media studies: "an online platform should be understood as a programmable digital architecture designed to organize interactions between users—not just end users but also corporate entities and public bodies." (Van Dijck et al. 2018)

A distinct example of web-based social machine is Wikipedia, where participants do not execute instructions, but are regulated by tight rules[8], and interact via a common software infrastructure to generate and maintain an encyclopaedia. News curation communities such as Reddit and Digg work in similar ways, centred around shared social practices and software infrastructures.

Yet a different example of social machines is given by the cases where humans do not need to be aware, they are participating. YouTube users generate high quality information about videos simply as a by-product of using the service (Covington et al., 2016). Similarly, eBay users participate in a very advanced mechanism to establish the price of goods, just by bidding for products.

The examples above show that there can be two fundamentally different ways to design a social machine, and we will call them 'design principles.'

In one case, participants are directly instructed to perform tasks chosen by others and rewarded for that. This covers the assembly line, the bureaucracy, and in the case of online social machines, might cover cases like employment of Mechanical Turk workers to join a more complex machine.

The other case is where humans choose to participate in an activity, e.g. editing Wikipedia, ranking items, watching a video, and, as the unintended result of their acts, the machine as a whole performs a task. Note that, in this case, the participants might not need to be willing or aware to be part of a mechanism. For example, von Ahn and Dabbish, (2004) found that in 1 month 5,000 people can provide high-quality annotation for more than 400,000,000 images just by playing a carefully designed sort of guessing game, called the 'ESP game'[9]. In general,

---

[8] There are specific guidelines, e.g., for creating content (e.g. editors should write in a neutral way, avoiding, understatement/overstatement, and self-published sources, etc), for reviewing articles (e.g. good article should meet certain standards, such as verifiability and broad coverage), or for solving disputes (e.g. editors can create a 'talk page' to discuss changes or controversial contents and request a third-party opinion where necessary). For a full list see : https://en.wikipedia.org/wiki/Wikipedia:List_of_policies_and_guidelines#Content_policies

[9] The ESP game consists of two players who are remotely looking at the same image and are rewarded for guessing which words the other use to describe it. Since they are randomly paired and unable to contact each other, it becomes a mind-reading game, hence it was called ESP (Extra Sensory Perception) game (von Ahn and Dabbish, 2004). The optimal strategy for each player was to enter the label that is most appropriate for the label within a certain amount of time. In a four-month experiment, van Ahn

the same mechanism is at work in gamified crowdsourcing apps, like FoldIt[10], where people perform some complex or costly tasks without realizing they are doing so.

In the first class, social machines are designed according to a top-down approach. Participants receive instructions specified by a designer and execute them. In the second class, social machines are designed according to a bottom-up approach. Participants voluntarily pursue their personal goals, but their individual actions and interactions are constrained in such a way that the resulting (i.e. emergent) collective behaviour turns out to be pursuing the goals of the overall machine. Ant colonies complete complex tasks without any of the ants being aware of that or being in control. Markets, auctions, online recommender and e-commerce systems, might be in a similar league.

In the top-down approach, the instructions might be specified in a high-level programme[11] and dispatched to the operating components possibly by means of a social compiler[12], a layer of the machinery that would break the programme into elementary operations and assign them to the parts (i.e. human participants and machines).

In the bottom-up approach, the system is best described in terms of mechanism design, i.e. as an emergent behaviour resulting from the interaction of multiple rational agents. When this happens, both participants and the whole machine are specified in terms of the goals they purse rather than the instructions they need to follow. Technically speaking, they behave like autonomous agents.

2.2 – Autonomous Social Machines

---

and Dabbish (2004) suggested that the ESP game could be used to tag all Google Images. In 2006 Google got the license to develop its own version of the game (Google Image Labeller).

[10] FoldIt (https://fold.it/portal/) is an online puzzle video game launched in 2008 by the University of Washington. By playing the game thousand users helped researchers to discover some protein configurations.

[11] Programming languages to specify and coordinate crowdsourcing workers already exist. An example is AutoMan, a "crowd-programming system" based on Scala which allows the programmer to manage some parameters (scheduling, budgeting and quality control) and to abstract the details of the tasks "so that human computation can be as easy to invoke as a conventional function" (Barowy et al. 2012, 641)

[12] A proposal to build "crowdsourcing compiler" has been put forward by Chen et. al. (2016). The crowdsource compiler would decide "which components of the task are best carried out by machine and which by human volunteers; whether the human volunteers should be incentivized by payment, recognition, or entertainment; how their contributions should be combined to solve the overall task; and so on." (Chen et al., 2016, 106)

A system is said to be autonomous when it acts under its own control, i.e. it is driven by some intrinsic goals[13]. Autonomous agents are called "goal-driven" or "rational" when they pursue goals, specified by a utility function that they try to maximise[14].

Autonomous agents can use information gathered from the environment to make their own decisions and control some aspect of their environment. They can be adaptive (capable of learning) and perform types of inference (capable of reasoning). We often use the language of belief, preferences and utility to describe them, but this is for convenience and not necessity (e.g. see Russell and Norvig, 2010 and Burr et al., 2018).

Biological organisms or even species can be described as autonomous goal-driven systems (maximising number of offspring), as well as engineered control-systems (e.g. a thermostat pursuing homeostasis), but also economic agents (e.g. maximising profit). But not all systems can be described in this way. For example, typical assembly lines would not be an autonomous system, as they would not be able to react to changes in the environment without new instructions coming from outside. Wikipedia might be a kind of hybrid case since, while editors are driven by some private interests, their individual and collective behaviour is influenced by a series of guidelines and norms which are continuously revised and debated by the participants of the community (i.e. they are not elicited by the system's utility function).

Social machines can be used to implement autonomous goal-driven agents, even if human participants are not aware of that. In this case, the participants should not be able to determine the behaviour of the whole system – otherwise the system would not be autonomous. We call them Autonomous Social Machines (hereafter, ASMs).

Examples of ASMs can readily be found over the Internet. They include the recommendation systems behind YouTube or Amazon, which could never perform their function if it was not for the structured activity of their users - which act as (unaware) participants while make use of the system for their own purposes. Indeed, those systems were created when certain products or users did not exist, yet they can autonomously process them appropriately, as they learn and generalise. Common users' actions (purchasing, filling a wish list, reviewing or searching items, rating a transaction, flagging a comment, etc.) translate into information and ultimately into recommendations (Ricci et al., 2011).

---

[13] Since autonomous agents are directed towards the attainment of some purpose ("telos" in ancient Greek), their behaviour might be called "teleological" and has been widely studied in Cybernetics (Wiener, 1948) and Artificial Intelligence (Russell and Norvig, 2010).

[14] In economics, the notion of 'utility' is a metrics representing a user's satisfaction derived from picking one among possible options (e.g. in a recommender system, options might be books or videos). We represent a user's ordering preferences over a set of alternatives by using a utility function.

The behaviour of an ASM is not dictated externally by any of its participants, nor is it pre-determined by its original designers: it is instead the emergent result of its interactions. So long as the human participants perform local tasks without controlling the system, the resulting social machine can be considered autonomous. While the goal of most recommender and marketplace systems is to increase either click-through rates or sales or profits (generally called interactions, or engagements, see Burr et al. 2018), none of the human participants has the same goals.

In the design of autonomous social machines, a crucial problem is to link effectively participant's utility with that of the whole system in a way that the pursuit of the former maximizes the latter. To solicit the desired behaviour from participants, the designer might need to devise a series of incentives which may be positive (monetary rewards, points, forms of social recognition, etc) or negative (fines, exclusion from participation, etc.). In so doing the designer will develop specific mechanisms which will try to influence participants' behaviour so as to make them act in a way that maximises the utility of the whole system.[15]

The study and the implementation of incentive schemes constitutes the main subject of mechanism design, a branch of game theory with broad applications in markets, auctions, voting procedures, etc. (e.g. see Börgers, 2015). Note that mechanism design is perfectly suited to O'Reilly request for clear outcomes – "the secret", he says, "is to identify key outcomes that we care about as a society" (O'Reilly, 2016, 293) – since it works backwards, i.e. it sets up goals before choosing the rules of the game. In this way, it is possible to solve "a centralized problem in an informationally decentralized system" (Naghizadeh and Liu, 2016).

---

[15] Note that similar mechanisms have been widely studied in nudging theories, where a typical problem is how to influence agents' behaviour by intervening on the presentation of available options (Thaler & Sustain, 2008). For a discussion of nudging in the context of AI see (Burr et al., 2018)

While (automated) incentive mechanisms are not necessary for the regulation of a social machine – e.g. assembly line workers compliance with the rules can well be enforced via traditional systems – their employment can make a difference in the development of algorithmic regulation. Not only they may create those conditions which promote compliance and facilitate implementation, as pointed out by O'Reilly, but can also shift the locus of power and control of the system to the locus of the reward function, as well as giving rise to issues that will be discussed later (see section n. 5), such as value alignment. But before exploring potential problems, we examine how an ASM can make adjustments based on the information produced by the participants and activate mechanisms of feedback-loop.

**3 – Reputation System and Credit Scoring**

The systems being proposed by O'Reilly to replace current governance methods are based on the (control-theoretic) notion of a feedback loop, so that individual actions or behaviours directly result into personal incentives. This is implemented by the intermediation of an infrastructure that keeps track of each participant, information relative to them, and their score (similar infrastructures are commonplace in the domains of Customer Relationship Management and Human Resources Management Systems).

The fundamental technology required for this kind of social regulation is akin to that of a Reputation System, a social machine evolved in online communities to process trust, combined with a more modern version of credit scores. It is also related to Performance Management Systems such as those used in warehouses of online shops, such as Amazon. If combined together, these quantities can form a score which functions as an incentive-system to foster a desired behaviour (e.g. trustworthiness or productiveness).

Reputation Systems arose in the context of web communities for promoting trust and good conduct among the group's members (Jøsang at al., 2005). They allowed users to rate each other after completing a transaction and aggregate these ratings to produce a reputation score (Jøsang at al., 2005). This enables a notion of trust in online environments where users have limited information about products and other users. Note that rating can work in slightly different ways: while quality ratings allow users to assess things like movies and restaurants, reputation rating allow members of a community to rate each other. For example, in TripAdvisor users rate hotels and restaurants, in IMDB they rate movies, in Yelp they review local businesses, etc. In eBay, Uber and AirBnB, however, users rate each other and, in so doing, they provide information about members' trustworthiness. In so doing, reputation systems offer an implementation of those social fictions that Harari describes as essential for the functioning of a large society (Harari, 2014)

Note that rating mechanisms can be included also in recommender systems[16]. However, the scoring mechanism plays a different role in recommendation and reputation. While in the former case the score is used to estimate the preferences of users, in the latter case it is used to enforce some standard, i.e. avoid bad service providers[17] and, for this reason, is said to function as a form of "collaborative sanctioning" (Jøsang at al., 2005), a feature well encapsulated in O'Reilly's proposal. In other words, if we consider transactions occurring between a supplier and a consumer, the same mechanism can be deployed either for the purpose of enabling consumers to make a more rational choice, or for encouraging suppliers to adopt a given behaviour.

Moreover, reputation systems can incorporate various mechanisms to avoid malicious attacks from unauthorised users, but also spamming, bias and distortion. Indeed, a reputation system may face the problem of participants looking for shortcuts or tricks which maximise their utility but do not align with the utility of the system. For example, a user may want to inflate others' perception about itself (e.g. to increase visibility) or not to report truthful information about others (Naghizahiadeh and Liu, 2013), or an Uber driver might want to refuse a short ride, in the name of his/her convenience but against the interest of the passengers. We might say that these systems have by now become one of the various immune systems for large online communities.

---

[16] Recommender systems that incorporate rating mechanisms could be based on collaborative filtering, a technique that make predictions of user's preferences based on similarity measures among users.

[17] In comparing Collaborative filtering systems with Reputation systems, Jøsang et al. (2005) observed that while collaborative filtering selects rating based on similarities among users (the premise is that users with similar tastes will prefer similar products), a reputation system assumes that all members should judge the quality of a service or product because the aim is "to sanction poor service" (Jøsang at al. 2005: 624). This connects to another important distinction: collaborative filtering assumes that all members are trustworthy whereas reputation systems suppose that some users will try to deceive for increasing their personal benefits (Jøsang at al. 2005: 625).

While reputation systems produce a score that reflects the subjective judgments of a group of individuals about another individual, product or businesses, other social machines can output a score that results from objective measurements of participants' actions. This may involve monitoring activities by means of sensors, cameras, smartphones, or more sophisticated tools[18], and generate a score that may capture relevant information about worker's performance, such as productivity or engagement[19].

In general, scoring systems could be derived from combining various signals: reputation, objective performance criteria and other attributes (e.g. demographics). In some cases, a social machine can combine multiple sources, such as reputation and performance metrics. For example, several on-demand apps, like Uber, Lyft or TaskRabbit employ both social feedbacks and work performance metrics, such as acceptance/cancellation rates and the number of tasks carried out, and both can contribute to derive penalties and rewards.[20]

In many countries a well-known example of scoring mechanism is a credit score. This is a number intended to represent the risk of lending money to a particular individual. Credit scores were first introduced by FICO decades ago in United States, but from 1989 there has been a general-purpose FICO score is used by credit card companies and banks, and other companies exist that provide similar services (we discuss how credit scoring regulate consumers' behaviour in section 4.3). Scores are based on data coming from consumer-credit files provided by "credit bureaus", which may include: payment history (35%, how fast you pay your bills, bankruptcies); amounts owed, credit utilisation (30%, how much you have borrowed out of your total allowance); length of credit history (15%), etc. [21]

---

[18] Think of Amazon's wristband which uses ultrasonic tracking and a haptic feedback system for controlling the worker's hands. The wristband vibrates against the worker's skin if the hand points to the wrong direction (Solon, 2018)

[19] The field dealing with the tracking of employee's behaviour to extract information about their performance is also called "people analytics". This include workplace technologies that analyse various signals such as emails' content, web-browsing patterns, list of apps opened and generate some score. For example, WorkSmart is a platform owned by Crossover (https://www.crossover.com/worksmart/#worksmart-productivity-tool) that takes photos of employees every 10 minutes, combines them with screenshots of their workstation and other data, and comes up with an "intensity score" (Solon, 2017).

[20] Penalties may include a temporary deactivation of worker's profile and fees, e.g. see Ticona et al. (2018). In Handy, for example the service professional agreement states that: "In the event a Service Professional's aggregate rating falls below the applicable minimum rating, Handy reserves the right to deactivate the Service Professional's access to the Handy Platform." (https://www.handy.com/pro_terms)

The effects of scoring systems, such as credit score and reputation systems, depend on the way they are used. If they determine how easy it is for users to receive a service (such as a loan, or a car ride), then any change in these scores directly affects the expected utility of users and so their existence is likely to influence user behaviour. In the sociological literature the same effect is called "reactivity": social measures, such as risk score or performance indicators, are called "reactive" because they "elicit responses from people who intervene in the objects they measure" (Espeland and Sauder, 2007, 2). For example, schools and universities have made relevant changes in reaction to being ranked. So, they have changed the way in which they select students, allocate resources and organise work in order to optimize their rank (Espeland and Sauder, 2007).

Likewise, the design of reputation mechanisms has direct consequences for the people being scored. For example, an Uber driver with a bad or poor reputation might be suspended from work and, as a consequence of this, be encouraged to improve his or her service. Reputation score has a clear impact also in eBay: the average sales price of the same item can increase by 3% for top-rated sellers (Xiang Hui et al., 2016)

The basic idea of using scoring mechanisms in algorithmic regulation is precisely to exploit reactivity to steer collective behaviour by using scores as incentives. This is where control theory meets social scoring.

**4 - Social Control and Feedback Loop: ASMs for Algorithmic Regulation**

So far, we have seen that there exist mechanisms that can monitor the behaviour of an individual and compute a score that captures how aligned that behaviour is with a given value function. That score can increase and decrease in response to behaviour or performance, just like a credit score does, and it can include elements that are typical of reputation systems. Generally, these mechanisms can be based on social machines like those described above and be autonomous (i.e. an ASM), in the sense that various participants can provide the signals that inform that score – either directly or indirectly – but no individual can significantly affect or control the system's outcome.

---

[21] FICO, also known as Fair Isaac Corporation, is an analytics software company based in California which was founded by William Fair and Earl Isaac in 1956. On FICO score and its components see also http://www.fico.com/25years/

ASMs which embed scoring mechanisms offer a natural setting for O'Reilly's proposal since they have a clear outcome, real-time measurements and mechanisms of adjustment (O'Reilly, 2013, 290), i.e. all the elements listed as necessary for a feedback-control system. In practice, they are capable of observing the state or actions of an individual, computing how it aligns with their goals, and then administering positive or negative incentives. So, their behaviour can be studied from the perspective of Control Theory.

4.1 – Regulation by control systems

A control system (the "controller") regulates the behaviour of another system (the "plant", or "controlled system") by taking actions that 1) depend on the current state of the controlled system and 2) affect it. To do this, the controller needs to a) observe the state of the controlled system, b) compare it with the target state (the "set-point"), and c) act on the controlled system to change its state. So, a controller must include sensors, actuators and a control algorithm. Clear goals (target state) and clear sensing (current state) are necessary, and the difference between these two states is used as a control signal[22].

O'Reilly's idea, of presenting citizens with incentives (rewards or punishments) that directly follow from their actions, aims at establishing a control loop - where citizens are assumed to act rationally, and therefore adapt their behaviour to maximise their utility. Actuators in this case are replaced by the capability to act on the score of an individual, i.e. to administer incentives. An important requirement, of course, is that the controller has clear targets, can predict (probabilistically) the consequences of its actions and can read the actual state of the controlled system.

In general, the behaviour of individuals can be influenced in various ways. There are forms of 'soft' control that increase the probability of an action being taken. Common methods may include nudging based on cognitive biases, or trading based on knowledge of economic incentives, and even extend to forms of coercion or deception (Burr et al. 2018). A mechanism of this type, when applied to entire societies, can potentially steer their collective behaviour and offer alternative methods to law enforcement.

---

[22] When the controller's action depends on such a difference (e.g. the house temperature and the temperature set on the thermostat) the process is said to be a "closed loop control system".

In reality, models of governance based on persuasive technologies already exist. Consider the problem of managing traffic: limiting driving speeds can be achieved by a system of laws and fines, or by offering some rewards. For example, Enschede (Netherland) has invested 36 million euros to deploy an app that creates personal mobility profiles and rewards good behaviour like cycling or walking (Naafs, 2018). Similarly, a form of (negative) incentive is the supplementary taxes which are added to the cost of unhealthy products (e.g. cigarettes or food and drinks with high level of fat) to discourage buyers and promote a healthier lifestyle.

In practice, however, the target-quantities observed by these regulatory systems may be replaced by proxies (or surrogates) that only partly align with the actual targets. For example, college and university rankings originated from the idea of publishing relevant indicators of the performance of higher education institutions, making them more accountable to the public (Espeland and Sauder, 2007). But, over the years, they turned into (objective) measures of prestige and a goal to be pursued for its own sake. The combination of multiple metrics, such as percentage of graduate students and the number of highly cited researchers, is in fact a construct that gives a limited understanding of what a university offers and, by the way, the result of a computation performed by a social machine.

In sociology this effect relates to the problem of "commensuration" (Espeland and Sauder, 2007), i.e. the practice of translating qualities into quantities. In particular, Espeland and Sauder (2007) pointed out that mechanisms of commensuration tend to change the focus of attention and sense making, i.e. "they reduce, simplify and integrate information" (Espeland and Sauder, 2007, 20)[23]. This mechanism has many important effects and a problematic one is the generation of unexpected and unintended reactions, which are often extraneous to the stated goal of the system – e.g. universities started hiring "ranking managers" and manipulating internal rules in order to increase their score (Espeland and Sauder, 2007).

---

[23] For example, rankings "are meant to simplify complicated information; they embody decisions that make vast amounts of other information, often qualitative knowledge that is hard to assimilate to this form, irrelevant" (Espeland and Sauder, 2007: 17).

However, whatever the effect, in a control system it is pointless to distinguish between its intended and unintended consequences since, from a cybernetic point of view, "the purpose of a system is what it does" (Beer, 2002). Therefore, to understand the functioning of a control system it is better to look at the observed effect rather than at the original intentions of the designer. If the system of credit scoring was initially designed to streamline bank decisions, but ended up changing consumer behaviour, then the credit score system has the effect (and therefore the purpose) of regulating consumer behaviour. In other words, the purpose of the system and that of its designer do not align. Likewise, even though university ranking systems were motivated by the need to increase accountability and support decisions of policy makers (e.g. how to distribute resources), the ranked institutions adapted their behaviour to the incentives (delivered by the score) and generated a number of second-order effects, which redefined the goal of the overall system (e.g. to promote standards of prestige and authority).

4.2 – ASMs for Social Control

If we take O'Reilly's proposal seriously, the key idea behind algorithmic regulation can take the form of an ASM with mechanisms of scoring for citizens and the resulting control loop. This would turn a society into a system where citizens are all automatically "enrolled" (not by opting into a private service, but as an essential part of their citizenship). The incentives that the system generates in terms of score would directly affect citizens' utility and opportunities, and their behaviour would adapt accordingly. The central question posed in this article relates to the possible social implications of such a turn.

This mechanism can be used in various ways and at different scales combining public and private resources, as O'Reilly suggests. For example, the city of San Francisco partnered with Yelp's restaurant review platform to share health inspection data and "create a safer, healthier dining experience"[24]. But, probably, the most ambitious project has been taken in China, with the creation of a Social Credit System. This is intended as an example, as other countries are active in the same space, with different projects.

---

[24] Each year the health department inspects local restaurant and assign a score based on how they respect health code regulation. In this way, the initiative aims at improving food safety and restaurant's attractiveness. See more here: https://www.codeforamerica.org/featured-stories/san-francisco-promotes-its-restaurant-inspection-data-on-yelp-to-improve-food-safety

China has been the first country to officially call for a unified system for social credit, where every single citizen is listed in a national database, and "social credit" information is appended to it. Different variants of that concept have been tried, and we have to distinguish the national-level system from various city-level experiments: the current proposals for the national system do not include a score, but rather the binary decision of adding a citizen to a black list or a red list on the basis of their behaviour, while certain city-level experiments (e.g. projects in Suining and Rongcheng), have explored the possibility of using actual scores and letter-grades, as have done some private companies (Larson, 2018; Creemers, 2018).

The system resulted from an official policy decision of the Chinese leadership to use technology-driven tools for social control, as a supplement to traditional forms of governance (Creemers, 2018). The idea originated in the context of financial credit – as a sort of Chinese version of FICO score – and expanded across domains to promote trust and honest conduct within society[25]. The plan, officially outlined in 2014, includes a timetable for the realisation of a Social Credit System by 2020, including five steps: "creating a legal and regulatory framework for the SCS, building credit investigation and oversight, fostering a flourishing market built on credit services, and completing incentive and punishment mechanisms." (Creemers, 2018, 12).

While the planning document does not refer to any scoring methods, in the county of Suining a local project put the idea in practice. Citizens were given 1000 credit points at the start, then points could be deducted for infringements of certain norms, e.g. drunk driving convictions (50 points), having a child without family planning permission (35 points), non-repayment of loans (30 to 50 points). Lost points could be recovered after a period of two to five years. On the basis of this score, citizens would be assigned to classes from A to D, and A-class citizens would have preferential access to employment opportunities, while lower-ranked citizens would face increased scrutiny in several areas, such as Party membership, enlistment in the military, loans, governmental support including basic social welfare (Creemers, 2018). The experiment, however, attracted criticism from State media, after which the A-D classification was dropped. A similar initiative, involving the assignment of a score and a ranking system with related punishments and rewards (Mistreanu, 2018), has been reported in Rongcheng and was listed among the main success stories for the propagation of social credit system[26].

---

[25] See the document: State Council. Planning Outline for the Construction of a Social Credit System (2014-2020). 14 June 2014. Available online: https://chinacopyrightandmedia.wordpress.com/2014/06/14/planning-outline-for-theconstruction-of-a-social-credit-system-2014-2020/

At present the Joint Punishment System is the main component which has been developed at the nationwide level. The system identifies a series of undesired behaviours (the so-called "trust-breaking acts"), which contribute to the creation of a blacklist, and associated punishments. Examples of untrustworthy acts include: endangering the personal health and life security of the popular masses, tax evasion, malicious evasion of debt, sales of fake and shoddy products, false advertising, gathering a mob to bring social order into chaos, refusing to carry out statutory duties, or gravely influencing the credibility of judicial bodies and administrative bodies, refusing to carry out national defence duties, refusing or evading military service, etc.[27]

Once a citizen is flagged with the "blacklist status", they can face restrictions in disparate fields: e.g. "They were barred from senior position in SOEs, financial sectors […] They were no longer allowed to travel first class, on high-speed trains, or on civil aircraft, to visit star-rated hotels or luxury restaurants, resorts, nightclubs and golf courses, to go on foreign holidays, to send their children to fee-paying schools, to purchase particular kinds of high-value insurance products, to buy or renovate their homes, or purchase cars." (Creemers, 2018:15).

As Creemers (2018) suggests, key elements in the design of the SCS are 1) a system that ensures a unique, lifelong identifier for each citizen and a register number for corporations; 2) a massive data infrastructure to collect information about subjects from different sources (e.g. bank, local authorities, social organizations, etc.); and 3) data mining technologies[28] to process the stored information to sort people and provide pathways for action.

---

[26] See the document "Notice concerning Issuing the Name List of the First Batch of Social Credit System Construction Demonstration Cities" of the National Development and reform Commission, and People's Bank of China, cited in footnote n. 81 in Creemers (2018)

[27] A more complete list can be found in document released by the State Council on 30th May 2016 (see note n. 1)

[28] However, as Creemers (2018) pointed out, at present there are few evidences on the employment of algorithmic analysis, apart from credit scoring: "As far as public documents indicate, anyone's social credit status will only be influenced by the history of their own conduct to the extent that it is covered through the SCS's remit. Technological analysis is used, however, to make mass information more manageable, accessible or technologically presentable." (Creemers, 2018, 22)

It is important to observe that similar components can be found in other national projects of smart governance. For example, Singapore's project started a wide process of digitalisation since the eighties and recently moved to the next steps, integrating Internet of Things (IoT) and AI solutions. Its ambition is to provide "anticipatory services", i.e. to solve issues before they are brought to the surface[29]. The project includes systems for detecting illegal parking, managing traffic and delivering services at key events of people's life[30].

In Europe, a model for smart governance is given by Estonia, where there are some analogies with the aforementioned cases: a unique identifier for each citizen linked to various repositories owned by private or public institutions (governments' departments, hospitals, pharmacies, schools, etc.), and tools for providing citizens and organisations with a vast array of services, from filing taxes, parking, buying tickets, to policing and voting (e-Governance Academy Foundation, 2016).

A distinctive feature of Estonia's project is that all data are not stored centrally, and a data exchange platform, X-Road, ensures secure communication among the various repositories. This reflects Estonia's most inherent philosophy about data: information should not be entered twice – Estonia's Public Information Act does not allow organisations to establish separate databases to collect the same data (Vassil, 2016) –, and individuals should be the owners of their data (e-Governance Academy Foundation, 2016; Heller, 2017). However, although the eGovernment ecosystem incorporates explicit legal principles for the protection of personal data, to the best of our knowledge, citizens' right to authorize or deny access to personal data is limited to the health-care system (Priisalu and Ottis, 2017) – e.g. does this apply also to education and business services? – and the possibility to opt-out of the ID card system is not allowed[31].

---

[29] https://www.zdnet.com/article/singapore-unveils-plan-in-push-to-become-smart-nation/ but also see the report "Digital Government Blueprint" https://www.tech.gov.sg/-/media/GovTech/DigitalTransformation/Digital-Government-Blueprint/dgb_booklet_june2018.pdf?la=en

[30] For example, in 2018 a new app was launched for starting families. Thus, new parents could be automatically informed about birth registration, infant care and kindergarten admissions (see: https://govinsider.asia/digital-gov/singapore-smart-nation-e-payments-national-digital-identity-anticipatory-services/)

[31] The ID card, which gives access to the services, are obligatory for all citizens, a rule that was approved in 2000 by the Estonian Parliament with the Identity Document Act (Vassil, 2016).

The article (Priisalu and Ottis, p. 445) implies that personal data is accessible to government officers, the only protection being an automatic audit trail that makes it possible to log any access to personal data.

At present Estonia government does not include any scoring or reputation mechanisms, but in future the system may evolve into a next generation of e-services as a result of the persistent data collection activities. For example, the use of predictive analytics has been advocated [32] in the context of disease prediction (e.g. type 2 diabetes).

Even though ASMs for nationwide social control do not exist in Europe, some argue that the seeds of a social scoring are already present in European countries. For example, the psychologist Gerd Gigerenzer observed that in Germany there is a credit score, called "Schufa", that assesses three-quarter of Germans and five million companies. He observed that people who aim to rent a house or get a loan in Germany are required to provide their "Schufa rating" (the analogous of FICO score in US) and "factors like "geo-scoring" can also lower your overall grade if you happen to live in a low-rent neighborhood, or even if a lot of your neighbors have bad credit ratings." (quoted in Jahberg, 2018).

4.3 - Examples from the Private Sector

Despite the recommendations of Tim O'Reilly, the governance systems currently being developed by states do not match those used in the private sector, but it is worth keeping an eye on that sector, to see what side-effects this management technology can have.

As we said an important area of application for algorithmic governance is credit scoring. The capability of regulating people's behaviour is a direct consequence of the huge impact that a credit score has on people's life: a bad credit score can result in higher interest rates for a loan or, even worse, a denial of loan application. But the information determining a credit score (i.e. the credit report) may in part affect other important decisions, such as hiring and renting – credit checks can in fact be requested also by employers, landlords and utility companies[33]. Of course, these effects translate into actions to be taken by the individual in order to avoid bad behaviours (e.g. delays in payments and losing income) that would negatively impact the score and create a bad credit reputation.

---

[32] See the interview to Kristjan Vassil (University of Tartu) on the next generation of e-services, available online: https://e-estonia.com/a-conversation-with-kristjan-vassil-on-the-next-generation-of-public-e-services/ and the claims of the Chief Information Officer Siim Sikkut quoted here: https://www.politico.eu/article/test-driving-the-ultimate-connected-society-e-stonia/

In future we may expect that the influence of credit score will expand beyond its original boundaries – for example some suggest application in online dating [34] - and a sign of this might be the growing numbers of tech companies issuing credit scores. For example, firms like Sesame credit (an affiliate of Alibaba), Tencent, and Lenddo issue credit scores also by making use of online shopping habits, social media and networking activities. Some argued that this phenomenon might have disruptive effects on how credit scoring regulates social behaviour (Gapper, 2018). So, while in traditional credit systems one improves its score by borrowing less and displaying self-control, social credit "tends to work in the opposite way — it gives users an incentive to buy and rent items through platforms, and to build a circle of active and highly rated friends." (Gapper, 2018)

A key example of algorithmic regulation in the private sector is Uber, a digital platform that coordinates nearly two million drivers[35] by means of an autonomous software agent. Rosenblatt (2018) summarises the situation of Uber drivers as working for an algorithm and says that: "the algorithmic manager seems to watch everything you do. Ride-hailing platforms track a variety of personalized statistics[36], including ride acceptance rates, cancellation rates, hours spent logged into the app and trips completed". She also adds that: "Uber uses the accelerometer in driver' phones along with GPS and gyroscope to give them safe-driving reports, tracking their performance in granular detail."  Rosenblatt (2018).

---

[33] For example, Experian, one of the main credit reporting agencies, declares that a job candidate may not be hired because of a bad credit report: and
https://www.experian.co.uk/consumer/guides/employment.html The same may also happen for ranting as reported by Equifax (another popular credit rating agency):
https://www.equifax.co.uk/resources/loans_and_credit/credit_checks_for_renting.html

[34] For example, Denyer (2016) claimed there are some online dating platforms which encourage users to "display their Sesame Credit scores to attract potential partners."

[35] If we consider other Uber services, such as Uber eat, riders are 75 million. See some facts and figures in Uber website: https://www.uber.com/en-GB/newsroom/company-info/

[36] The assessment of drivers combines a star rating system, cancellation rate, acceptance rate and safety (e.g. through GPS tracking). Each of these can contribute to quality assessment, for example: "Cancellations create a poor rider experience and negatively affect other drivers". See more on Uber Community Guidelines: https://www.uber.com/legal/community-guidelines/us-en/

Notifications on performance and incentives are delivered on a regular basis to each driver. For example, Rosenblatt (2018) reports that drivers are shown selected statistics as motivating tools (e.g. "you are in the top 10% of partners") and are informed on the areas in high demand.

The rating systems can also contribute to the deactivation of drivers: "In certain services on Uber's platform, if drivers fall below 4.6 stars on a 5-star rating system, they may be "deactivated" — never fired. So, some drivers tolerate bad passenger behaviour rather than risking retaliatory reviews" (Rosenblatt, 2018).

In 2015 an article in Forbes already described this situation based on a blog post[37] of Silicon Valley CEO Peter Reinhardt. The article stresses that the main effect of labour platforms following a Uber-like model is that of "replacing middle management with APIs" (Kosner, 2015). This is described as a trend that will divide jobs into two categories, those below the API, managed by a software platform, and those above the API, in charge of making or controlling such platform (Kosner, 2015).

The use of metrics and algorithms to manage a working environment has also been used in Amazon warehouses, as reported by New York Times in 2015 (Kantor and Streitfeld, 2015). This article describes warehouses where workers are monitored by electronic systems to track performance, and office workers can report on each other's performance through a tool called "Anytime Feedback", which provides part of the metrics that rank workers – according to Kantor and Streitfeld (2015) the bottom of the ranking is eliminated periodically.

**5 – Discussion**

The governance of increasingly large organisations or entire societies has so far relied on a set of methods and principles that evolved over the centuries, such as representative democracy and law enforcement. As we see various proposals for the introduction of new technologies for social regulation, we should be clear that this turn would involve a transfer of power from current institutions to new ones, and that this transfer might not easily be reversible. Therefore, a careful examination of positive and negative consequences would be essential, as well as a transparent public debate including all parts involved, before - not after - any deployment.

---

[37] The blog post entitled "replacing middle management with an API" is available online: https://rein.pk/replacing-middle-management-with-apis

The stated benefits of various forms of digital governance, with or without predictive analytics or incentive systems, have been mentioned above: increased transparency and efficiency (from the Estonian project), increased compliance and morality (from the Chinese projects), faster adaptation and control (from various commercial projects). Moreover, algorithmic regulation may offer greater flexibility and reduce the workload of regulators with saving of costs and time, i.e. it would allow to "govern least" (O'Reilly, 2013)

In the subsections below, we focus on possible consequences of adopting algorithmic regulation of society, and at the end we also describe one possible way in which this might emerge without being explicitly adopted. Questions of various orders should be addressed as a matter of urgency, by different sectors of society and academia. These might regard issues of stability (*How do we deal with undesirable dynamics, such as wealth concentration and low social mobility?*), personal autonomy (*how can we prevent an ASM from bypassing human deliberation?*), value judgments (*to what extent could an ASM elicit change of habits?*), and power (*could an ASM redefine social ties and social norms?).*
While we mention some of them below, we do not claim that we can cover all of them: this new area of technological and social change does require urgent multidisciplinary attention. The purpose of this article is to pose the question of what implications this technology might have for society, multiple voices will be needed to address it.

The concerns of possible problematic consequences can be divided into three levels: technical, ethical and political. By this we mean to separate risks that would result from an imperfect technical realisation of the project, from those that would result from the concept itself of algorithmic control of society.

5.1 Technical level

While the engineering aspects of building an ASM can be solved by using infrastructures and technologies of the sort used today by online companies (e.g. Facebook) or banks, the emerging effects of connecting multiple interacting parts are not explored. Once everything is connected with everything, higher order interactions can emerge. If the opportunity to have a job relates to online purchases or reading habits, for example, a new unexplored interaction is created. So, creating multiple separate feedback loops and scores might be safer and more stable than merging them and combining them into a general national scoring system. The experience of filter bubbles, public-opinion manipulations, market flash crashes, should be kept very much in mind.

As the individual scores both affect the behaviour of citizens and are affected by it, there is the potential for feedback loops. If we also introduce reputation into the equation, then feedback loops can lead to stigmatisation and discrimination (positive feedback loops amplify small differences). If we use social connections of a citizen as one of the signals to compute their social credit, we might automatically create the potential for stigmatisation of low-rank people, which would create a dynamic of rich-gets-richer and poor-gets-poorer, or self-fulfilling prophecies: if people believe that they will be penalised by associating with low-scoring people, then they will adapt their behaviour – thereby reducing the opportunities of low-scoring people.

One should not exclude that similar dynamics might end up generating a power-law distribution, like those observed in the disproportionate distribution of wealth (Jha, 2011). In a social scoring system a similar distribution would create an elite of people clustering around the first positions and the vast majority variously distributed on the rest, so that only a small portion of the population would account for the average score of a country's citizens.

Social mobility would be a connected issue, if negative feedback blocks people into the same rank with little chances of moving. The opposite can also be problematic, with excessive volatility, due to positive feedback loops. Either way, the design of these systems might directly affect society structure and should not be taken lightly.

Instabilities might potentially also lead to flash-crashes, or inflation, as well as spontaneous growth in inequality. Just like filter bubbles and market flash-crashes, feedback effects should be expected here, and remedies should be planned ahead. *What kind of dynamics can we expect? What lessons might be learnt from analogous complex interactions (e.g. algorithmic pricing) and applied to an ASM? What are the risks of instability?*

The purpose of an autonomous system is defined by its value function, and the control system centred around the social score is no exception: it will incentivise specific behaviours in the user and in society as a whole. The problem is that we typically can only measure the state of the controlled system (society) through proxies, that we assume to be well aligned with our actual goals. The quality of that approximation may change with time, as the controlled system evolves, and over time we might have a system that is actively encouraging the pursuit of behaviours that do not align with the original goals. Citizens may have to follow behaviours that are actually sub-optimal or negative, in order not to be disadvantaged.

Furthermore, if the value function is itself relative to the rest of society (e.g. bonus is given to the top-10% in the group, or for moving towards the mean-behaviour in the community, etc.) then this can create not only competition, but also a *drifting value* function. Rational citizens might then work against their own long-term good, locked into this behaviour by the automated system of incentives. In academia – to cite a familiar example to our readers – this has led many to prioritise publication numbers over quality, and ultimately to a multiplication of academic journals only justified by the spurious identification of publication rates with scientific productivity. *In a similar scenario who will be in charge of changing the overall value function? what kind of safeguard measures do we need to avoid undesired drift, or worst, detrimental effects?*

5.2 Cultural and ethical level

The idea of using feedback loops for the control of social systems is as old as cybernetics. Stafford Beer pioneered the use of cybernetics in management, introducing the notion of "total system" (Beer, 1975), and the founder himself of the field, Norbert Wiener, devoted a book to the topic (Wiener, 1950). In legal scholarship, a similar view underlies the analysis of decentred regulation whose essential elements include, among other things, complexity, fragmentation of knowledge and the exercise of power, and interdependencies (Black, 2002). Still, with his popular motto "the code is the law", Lawrence Lessing suggested how the architecture of the Net (and its possible evolutions) can become a perfect tool of social control (Lessing, 2006)[38].

---

[38] "the invisible hand of cyberspace is building an architecture that is quite the opposite of its architecture at its birth. This invisible hand, pushed by government and by commerce, is constructing an architecture that will perfect control and make highly efficient regulation possible" (Lessing, 2006: 19)

So, what is proposed by O'Reilly, and is being explored in China and other countries, is not an innovation *per se* – the novelty is that we now have the technical means to gather individual information and administer individual incentives on a vast scale.

This article suggests that any such system requires the creation of a powerful intermediator: the infrastructure that gathers and manages individual information, computes scores and incentives, and organises the individual scores. Whoever sets the rules for that calculation has the power to steer the social group represented in the database. This observation, which is inherent to the design of an autonomous agent, suggests a possible cultural fallacy in the ideal of a decentred exercise of power. Fragmentation of knowledge and distributed information processing might give the participants only the illusion of freedom since their choices would be continuously coordinated and influenced by the controlling agent (i.e. the system-level utility function).

Even though the idea of self-regulation is fascinating because of its capability of harnessing diffused power and knowledge within society, the regulation operated by an ASM would be anything but neutral[39]. The introduction of intelligent machines at the centre of our society creates an intermediator with enormous privilege: it would behave as the best-informed player at each level of social interactions and the coordinator of all partial decisions (Wiener, 1950).

The transition towards a non-neutral ASM regulating a group of people (be that an online community or a nation) creates a number of interrelated issues concerning human autonomy and freedom. These issues may arise from bypassing participants in making decisions and taking various assumptions for granted in the design process. For example, the idea of moving the regulation of domestic work into a platform like Handy or TaskRabbit, supposes that every participant agrees on various points (i.e. the rules of the game): e.g. that everyone has a device to access the service and knows how to use it; that customers' ratings determine the salary of the taskers; that members communicate only by using the platform, and so forth.

---

[39] As Gillespie observed, the ability to orchestrate interaction among users is often contrasted by claims of impartiality made by the owners of media platforms: "from their earliest presentation they have often characterized themselves as open to all comers; in their promotion they often suggest that they merely facilitate public expression, that they are impartial and hands-off hosts, with an "information will be free" ethos, and that being so is central to their mission" (Gillespie, 2017 p. 4)

In some cases, the implicit acceptance of the rules of the game as a precondition to opt-in might be not problematic (e.g. eBay or Airbnb) but in others decisions should be negotiated with all members of the community. This is the case of services involving citizens and the possibility to coerce them by imposing conditions they might not share or agree with. For example, the decision of using a digital infrastructure to regulate the activity of a school would require a process of reflection and discussion at different levels (individual, family, the board of teachers, etc.) to enable everyone to express their own opinions and participate in the deliberative process. In fact, a choice of that kind would not be just a technical change but also a policy decision premised on various assumptions (e.g. do all families have access to digital devices? do all agree on the educational values supporting certain functionalities?). One of the problems posed by an ASM regulating society is that it tends to obfuscate what decisions its participants are entitled to and what could be made without their involvement (i.e. without any infringement of citizens' rights). So, fundamental questions for policy-makers, designers and scholars, are: *what decisions and assumptions does the system take for granted? how may these impact the life of people? what decisions should be debated by all members of the community? And, more in general, in which ways could an ASM change democratic participation and social inclusion?* Answering these questions involves a perspective of cooperative responsibility as an ASM results from the dynamic interaction of different types of actors (the participants, the owners of the platforms, etc.) and requires developing a process of public deliberation (Helberger et al., 2018)

An immediate dilemma emerging from this practice would be: should the scores be public? should they be shared with the individuals? Should the various metrics and formulae be public? Each answer would lead to different problems: complete transparency would create the risk of people 'gaming' the system, less than that would reduce people's chances to calculate their best course of action.

Another set of problematic issues regards the *internalisation* of the quantities which may qualify participants in a social machine (e.g. their reputation or their performance). When we introduce a new intermediator to manage social regulation, we may internalise and socialise quantities such as our social score, effectively self-regulating us according to the rules of the system.

The application of numbers to qualify people's life not only marks a departure from modern conceptions of human dignity[40], but also promotes a new form of moral exemplarity. The computation of scores has in fact the function of producing appropriate incentives and, with them, a desired change of (external) behaviour. But, as well as that, it may also involve a change of participants' cognition (Espeland and Sauder, 2007), i.e. how do they interpret and value those numbers. As participants assimilate connections between scores and benefits/penalties, they may internalise authoritative standards against which assess themselves and others. For example, connecting a credit score to the ability of finding a partner invites people to value creditworthiness as a proxy of attractiveness, and making public high-rated users (e.g. eBay) or high-scored citizens (e.g. China) are effective ways to create role models and inspiring examples. Can we imagine a new class system based on these scores? How powerful would be whoever can affect that class system?

We should also note here that recent work (Burr and Cristianini, 2019) suggests that very personal information, such as beliefs, attitudes and aptitudes, can be inferred from the analysis of online behaviour, which poses new issues related to freedom of expression, as well as to privacy and fairness.

As an ASM expands its ability to implicitly convey values and norms, one might ask what the role of free will in participants' reaction is, and to what extent their change of beliefs and values is in fact due to persuasive mechanisms (see also Burr et al, 2018) rather than intentional actions. The development of moral habits requires some degree of automatism - it is a common idea in both ancient and contemporary philosophers that habits requires repetition and routine movements (e.g. see Aristotle and William James) - but this is not just a passive exercise. The acquisition of habits needs also the faculty of judgment deliberating what are good and bad habits (see e.g. the role of practical rationality/wisdom in Aristotle or Dewey's value judgments). *To what extent could an ASM elicit change of habits? How can participants exercise moral judgment in a social machine which tries to continually adjust their behaviour and has the potential to change their beliefs and values? How do we help participants purse their goals in the respect of their freedom, autonomy and dignity?*

5.3 Political and social level

---

[40] In Western culture moral principles and human rights rest on Kantian conception of human dignity considered as "an intrinsic worth" that is distinguished from something that has a price and that can be exchanged for something else (Kant, 1785).

Attention should also be paid to the political and social implications of this technology. One political concern that has been put forward, is that the emergence of a form of governance is strongly related to technological innovation, and intelligent systems may contribute to the creation of new authoritarian regimes (Harari, 2018). In this scenario, the new face of dictatorship might not necessary be that of an unpleasant, authoritarian ruler, but that of an efficient autonomous system which exploits the information produced by citizens with the aim of prescribing the behaviours that best maximise its specific goal (be that "trust", "honesty", "excellence", "transparency", "accountability", etc). Using the terminology of this article we could consider that the next generation of dictators might look like an ASM, i.e. an autonomous social machine that mediates every interaction and, thanks to the collected information, computes scores and delivers incentives to influence participants' behaviour.

While the example of the Chinese system is not an ASM - as we said, the apparatus rests on a binary decision (not a score) and incentives are not generated automatically by an algorithm -, the system proposed by O'Reilly would definitely be. The technical properties we surveyed in the previous sections, now invite us to reflect on their social impact.

As the deployment of social-regulation technologies is essentially an exercise in reforming the way we control society, it is at its core a problem of power. This should be analysed with the conceptual tools relative to power and institutions, which is too large a topic to be discussed here. However, we will discuss here a few basic considerations with respect to contemporary conceptualizations of power.

In social and political science, the notion of power refers to a form of control that is exercised on a group of people by some authority (e.g. a government, a monarch, a religious leader, a manager, etc.). Its exercise may involve the use of physical force, but modern societies have also developed non-violent methods to control populations. These include a set of mechanisms based on scientific knowledge and technical innovations to manage effectively social organisations (e.g. cities, hospitals, schools, prisons, factories, etc.), an approach that Michel Foucault calls "biopower" (Foucault, 2007). According to Foucault modern institutions, starting from the military system, have learnt such methods and developed common characteristics that we might find also in a social machine: hierarchical surveillance - spaces and hierarchies are designed in a way to make "people docile and knowable" (Foucault, 1991, 172) - , mechanisms of gratification and punishment which correct and differentiate individuals (the so-called "normalizing judgments"[41]), and perpetual forms of examination which define the particular status of each individual.[42]

It would be interesting to analyse how ASMs fit within the general progression of methods of social control described by Foucault, and we leave this as an important open question (Foucault, 1991).

More generally, future work should attempt to answer the following questions; *what kind of society could an ASM produce? what social norms would it reflect? Who would be responsible for their embedding into the system? How do we guarantee the pluralism of values in a system that might tend to homogenize individuals?*

In political philosophy, the framing of social bonds in purely utilitarian terms would connect to the notion of "private society", i.e. a group of individuals whose "motivational horizons do not extend beyond the people and projects that are the focus of their personal lives" (Waheed, 2018). A society of this type tends to undermine the idea of common good[43] because each individual would care about only those goods with a direct impact on its private life. Even in a context of an ASM, where incentives might be deployed for promoting a specific common good, say equal access to opportunities, some philosophers would argue that this would not be

---

[41] "The art of punishing, in the regime of disciplinary power, is aimed neither at expiation, nor even precisely at repression. It brings five quite distinct operations into play: it refers individual actions to a whole that is at once a field of comparison, a space of differentiation and the principle of a rule to be followed. It differentiates individuals from one another, in terms of the following overall rule: that the rule be made to function as a minimal threshold, as an average to be respected or as an optimum towards which one must move. ft measures in quantitative terms and hierarchizes in terms of value the abilities, the level, the 'nature' of individuals. It introduces, through this 'value-giving' measure, the constraint of a conformity that must be achieved. Lastly, it traces the limit that will define difference in relation to all other differences, the external frontier of the abnormal" (Foucault, 1991,182)

[42] Foucault took Bentham's panopticon as the best representation of a complex of mechanisms governing a social body: "a generalizable model of functioning; a way of defining power relations in terms of the everyday life of men" (Foucault, 1991, 205)

[43] The notion of common good is controversial and different accounts exists. In this article we refer to the common good as the set of interests that all members of a community care about in virtue of their mutual relationships, such as civil liberties and public safety (Waheed, 2018).

enough: even in that case individuals would act for the sake of some private benefit and fail to perceive a "relational obligation towards common affairs"[44] (Waheed, 2018)

These intuitions suggest that a society implementing algorithmic regulation may risk to redefine social relations: individuals might be encouraged to establish strategic relationships (seek people that would increase social score and avoid those that would penalize it) according to the incentives set up by the controlling agent. *What would be the future of the common good in a society regulated by an ASM? How could an ASM contribute to protect common interests which would stand independently of individual agents' interest (be that of the controlled or the controlling agent) and, potentially, in contrast to anyone of them?*

**Conclusions**

The main purpose of this article is to pose a series of urgent questions, rather than answering them. Algorithmic regulation in this moment is little more than a tempting idea in academic, policy and entrepreneurial circles, but many of its components already exist, and furthermore many recent developments suggest that there is political interest in some version of it, although by other names. While it might be unlikely that in Europe, we would see an explicit effort to fully deploy this kind of system, there is still the risk that this might emerge on its own, as the result of mergers and drifts of related systems, much like it seems to be happening with national DNA or ID systems or surveillance

The risk of drifting into some version of algorithmic regulation is real. For example, it may start from a specific or local project, then grow by increasingly incorporating new capabilities (e.g. linkage of information, inference mechanism, classification systems, etc). Maybe they could start as an opt-in system, but then by *gravitational-pull* end up being unavoidable, and *de facto* mandatory, as is today being online. The most likely starting points for this drift are either a national ID system, acting as the scaffold to connect various sources of personal information, or a scoring system for specific categories (e.g. for certain professions) with the ability of growing to cover an increasingly large domain of society.

---

[44] These relational obligations are analogous to those found in family: parents are required not only to feed and clothe their children, they are also required to care about them (Waheed, 2018), e.g. their education, their mental and physical health, their moral development, etc. Something analogous exists also for civic relations.

Any social platform where users interact will end up influencing their behaviour, and not in a neutral way. An instructive example of unintended drift – that would be familiar to the readers of this article – is that of ORCID numbers, initially introduced as a way to solve homonymy among researchers, which however nudge all members to accept various levels of service, linking their name to citation indices, and essentially nudging towards a scoring system for academics. Since many journals demand that their authors have an ORCID number, there is a clear path a hypothetical moment where all researchers would have to accept the scoring system chosen and therefore act accordingly. Of course, there is nothing particularly sinister in academic scoring, this example is only intended to illustrate how drift can happen.

This simple consideration adds one more question to the list of problems that we encourage our colleagues to work on: that relative to the opting-out / opting-in dilemma, which keeps on emerging in different areas of digital ethics.

We call "*gravitational pull*" the problem posed by technologies (e.g. algorithmic regulation) when they exert a force that brings ever larger portions of people's life into them. As the system scales up, the cost of opting-out increases with the size (or coverage) of such a system. Not only is this used in viral marketing strategies, but this also creates a Nash equilibrium where everyone is part of the system: at that point there is significant cost for each individual to leave. Could a business today afford not being on the internet? Is it still meaningful to claim that people have freely opted into such a system?

In the case of social scoring, imagine the situation where if a private entity takes the role of endorsing people, and gathers significant following, such that citizens are at a disadvantage by not opting-in, what obligations does that business have? Are these the obligations that currently apply to (financial) credit scoring systems? Would that private actor have the right to demote or expel (i.e., excommunicate) an individual? Of course, these issues become even more pressing if the endorsement is operated by a public institution.

As we said, the problem of "*gravitational pull*" relates to that of "*opting-out*": what do you do when most other members of your community agree to be scored, to quantify their level of trustworthiness? If you join, you accept the scoring rules and their consequences, as well giving increased coverage to the system. If you do not join, you might be stigmatised, potentially losing access to opportunities.

This is just an example of the several dilemmas that emerge from the study of persuasive technologies and algorithmic regulation, and they all - eventually - merge at the same place: the need to give a new, fresh and multidisciplinary look at the issue of autonomy and social ties, in the new situation where technology brings new challenges to that fragile concept. We hope that scholars in Ethics, Sociology and Engineering will find a way to jointly address that question.

**Acknowledgments**.

NC and TS were supported by the ERC Advanced Grant ThinkBIG.

Smart, P. R. and Shadbolt, N. R. 2014. Social machines. Ed. M. Khosrow-Pour, *Encyclopedia of Information Science and Technology.* Hershey, Pennsylvania: IGI Global

Solon, O. 2017. Big Brother isn't just watching: workplace surveillance can track your every move. The Guardian. November, 6. Accessed 11 July 2018
https://www.theguardian.com/world/2017/nov/06/workplace-surveillance-big-brother-technology.

Solon, O. 2018. Amazon patents wristband that track warehouse workers' movements, *The Guardian*. January, 31. Accessed 2 June 2018
https://www.theguardian.com/technology/2018/jan/31/amazon-warehouse-wristband-tracking

Sullivan P (2018). Life Insurance Offering More Incentive to Live Longer, *The New York Times*. September, 19. Accessed September 20, 2018
https://www.nytimes.com/2018/09/19/your-money/john-hancock-vitality-life-insurance.html

Thaler, R., and Sunstein, C. 2008. *Nudge: improving decisions about health, wealth, and happiness*. London: Yale University Press.

Ticona, J., Mateescu, A., and Rosenblat, A. 2018. Beyond Disruption. How Tech Shapes Labour Across Domestic Work and Ride hailing, *Data & Society Report*

Van Dijck, J., T. Poell, and de Waal, M. 2018. *The Platform Society. Public values in a connective world*. Oxford: Oxford University Press

Vassil, K. 2016. Estonian E-Government Ecosystem: Foundation, Applications, Outcomes. Background paper for the World Development Report 2016*,* World Bank, Washington DC

von Ahn, L. and Dabbish, L. 2004. Labeling images with a computer game. In Proceedings of the *SIGCHI Conference on Human Factors in Computing Systems (CHI '04)*, 319-326
http://dx.doi.org/10.1145/985692.985733

Waheed, H. 2018. The Common Good, *The Stanford Encyclopedia of Philosophy (*Spring 2018 Edition), Ed. E. N. Zalta, Accessed 15 October 2018
https://plato.stanford.edu/archives/spr2018/entries/common-good/

Weber, M. 1978. *Economy and Society*. Los Angeles: University of California Press

Wiener, N. 1948. *Cybernetics, or control and communication in the animal and the machine*. Cambridge, MA: The MIT Press

Wiener, N. 1954*. The Human Use of Human Being*. Boston: Da Capo Press

Xiang Hui, M.S., Zequian, S. and Neel, S. 2016. Reputations and Regulations: evidence from eBay. *Management Science* 62(12): 3604–3616

Yeung, K. 2017. Algorithmic regulation: a critical interrogation. *Regul. Gov.* doi:10.1111/rego. 12158
39